\documentclass[]{spie}  

\usepackage{amsmath,amsfonts,amssymb}
\usepackage{graphicx}
\usepackage[colorlinks=true, allcolors=blue]{hyperref}

\title{Large Observatory for x-ray Timing (LOFT-P): A Probe-class Mission Concept Study}

\author[a]{Colleen A. Wilson-Hodge}
\author[b]{Paul S. Ray}
\author[c]{Deepto Chakrabarty}
\author[d,e]{Marco Feroci}

\author[f]{Laura Alvarez}
\author[a]{Michael Baysinger}
\author[a]{Chris Becker}
\author[g]{Enrico Bozzo}
\author[h]{Soren Brandt}
\author[a]{Billy Carson}
\author[a]{Jack Chapman}
\author[a]{Alexandra Dominguez}
\author[a]{Leo Fabisinski}
\author[a]{Bert Gangl}
\author[a]{Jay Garcia}
\author[i]{Christopher Griffith}
\author[f]{Margarita Hernanz}
\author[a]{Robert Hickman}
\author[a]{Randall Hopkins}
\author[a]{Michelle Hui}
\author[a]{Luster Ingram}
\author[j]{Peter Jenke}
\author[k]{Seppo Korpela}
\author[l]{Tom Maccarone}
\author[m]{Malgorzata Michalska}
\author[n]{Martin Pohl}
\author[o]{Andrea Santangelo}
\author[p]{Stephane Schanne}
\author[a]{Andrew Schnell}
\author[q]{Luigi Stella}
\author[r]{Michiel van der Klis}
\author[r]{Anna Watts}
\author[s]{Berend Winter}
\author[s]{Silvia Zane}
\author{on behalf of the LOFT Consortium, the US-LOFT SWG, and the LOFT-P collaboration}

\affil[a]{NASA Marshall Space Flight Center, Huntsville, AL, USA}
\affil[b]{U.S. Naval Research Laboratory, Washington, DC, USA}
\affil[c]{MIT Kavli Institute for Astrophysics and Space Research, Cambridge, MA, USA}
\affil[d]{INAF-IASF, Rome, Italy}
\affil[e]{INFN Roma Tor Vergata, Rome, Italy}
\affil[f]{ICE (CSIC-IEEC), Barcelona, Spain}
\affil[g]{ISDC, Geneva, Switzerland}
\affil[h]{DTU, Kongens Lyngby, Denmark}
\affil[i]{NRC Research Associate, resident at U.S. Naval Research Laboratory, Washington, DC, USA}
\affil[j]{University of Alabama in Huntsville, Huntsville, AL, USA}
\affil[k]{University of Helsinki, Helsinki, Finland}
\affil[l]{Texas Tech University, Lubbock, TX, USA}
\affil[m]{Space Research Centre, Warsaw, Poland}
\affil[n]{DPNC, Geneva, Switzerland}
\affil[o]{Tuebingen Univ., Tuebingen, Germany}
\affil[p]{IRFU, CEA Saclay, France}
\affil[q]{INAF-OA, Rome, Italy}
\affil[r]{Univ. of Amsterdam, Amsterdam, Netherlands}
\affil[s]{Mullard Space Science Laboratory, University College London, London, UK}
\authorinfo{Further author information: (Send correspondence to C.A.W-H.) \\
C.A.W.-H. E-mail: colleen.wilson@nasa.gov, Telephone: 256 961 7624 \\
LOFT Consortium \url{http://www.isdc.unige.ch/loft/index.php/loft-team/consortium-members} \\
LOFT-P Collaboration \url{https://sites.google.com/site/loftpmissionpage/home}}

\begin{document} 
\maketitle
\begin{abstract}
LOFT-P is a mission concept for a NASA Astrophysics Probe-Class ($<$\$1B) X-ray timing mission, based on the LOFT M-class concept originally proposed to ESA’s M3 and M4 calls. LOFT-P requires very large collecting area, high time resolution, good spectral resolution, broad-band spectral coverage (2--30 keV), highly flexible scheduling, and an ability to detect and respond promptly to time-critical targets of opportunity. It addresses science questions such as: What is the equation of state of ultra dense matter? What are the effects of strong gravity on matter spiraling into black holes? It would be optimized for sub-millisecond timing of bright Galactic X-ray sources including X-ray bursters, black hole binaries, and magnetars to study phenomena at the natural timescales of neutron star surfaces and black hole event horizons and to measure mass and spin of black holes. These measurements are synergistic to imaging and high-resolution spectroscopy instruments, addressing much smaller distance scales than are possible without very long baseline X-ray interferometry, and using complementary techniques to address the geometry and dynamics of emission regions. LOFT-P would have an effective area of $>$6 m$^2$, $>10\times$ that of the highly successful Rossi X-ray Timing Explorer (RXTE). A sky monitor (2--50 keV) acts as a trigger for pointed observations, providing high duty cycle, high time resolution monitoring of the X-ray sky with $\sim$20 times the sensitivity of the RXTE All-Sky Monitor, enabling multi-wavelength and multi-messenger studies. A probe-class mission concept would employ lightweight collimator technology and large-area solid-state detectors, segmented into pixels or strips, technologies which have been recently greatly advanced during the ESA M3 Phase A study of LOFT. Given the large community interested in LOFT ($>$800 supporters\footnote{\url{http:// www.isdc.unige.ch/loft/index.php/loft-team/community-members}}, the scientific productivity of this mission is expected to be very high, similar
to or greater than RXTE ($\sim 2000$ refereed publications). We describe the results of a study, recently completed by the MSFC Advanced Concepts Office, that demonstrates that such a mission is feasible within a NASA probe-class mission budget. 
\end{abstract}


\section{INTRODUCTION}
\label{sec:intro}  
\medskip
\textit{LOFT-P} is a probe-class X-ray observatory designed to work in the
2--30 keV band with huge collecting area ($>10 \times$ NASA's highly
successful \textit{Rossi X-ray Timing Explorer (RXTE)}) and good spectral
resolution ($<$260~eV). It is optimized for the study of matter in
the most extreme conditions found in the Universe and addresses
several key science areas including:
\begin{itemize}
\itemsep 0in
\topsep 0in
\parskip 0in
\partopsep 0in
\item{Probing the behavior of matter spiraling into black holes (BHs) to
  explore the effects of strong gravity and measure the masses and
  spins of BHs.}
\item{Using multiple neutron stars (NSs) to measure the ultradense
  matter equation of state over an extended range.} 
\item{Continuously surveying the dynamic X-ray sky with a large duty
  cycle and high time-resolution to characterize the behavior of X-ray
  sources over a vast range of time scales.}
\item{Enabling multiwavelength and multi-messenger study of the
  dynamic sky through cross-correlation with high-cadence time-domain
  surveys in the optical and radio (LSST, LOFAR, SKA pathfinders) and
  with gravitational wave interferometers like LIGO and VIRGO.}
\end{itemize}
Detailed
simulations\cite{yellowbook, Watts2016} have
demonstrated that an order of magnitude larger 
collecting area than \textit{RXTE} (i.e., $>$6 m$^2$) is required to
meet these BH and NS objectives, and a previous engineering 
study\cite{Ray2011}
has shown that such an instrument is too large for the Explorer (EX)
class and requires a probe-class mission.

\begin{figure} [ht]
   \begin{center}
   \includegraphics[height=5cm]{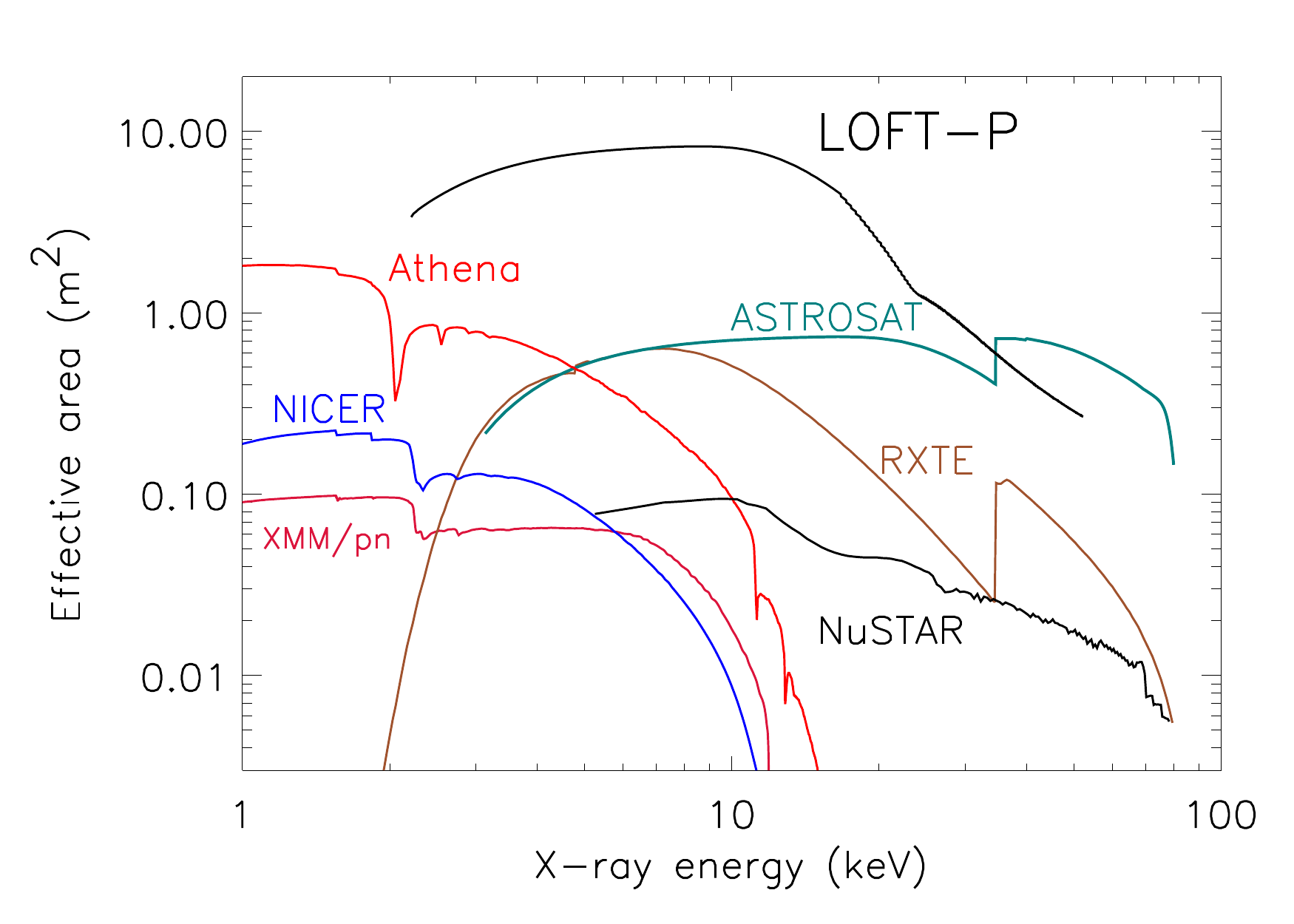}
   \end{center}
   \caption{ \label{fig:loft-p_effarea} Effective area as a function of area shown for the LOFT-P LAD baseline concept. Several existing and planned missions are shown for comparison.}
   \end{figure}

The \textit{LOFT-P} mission concept, which has been under study in
both the Europe and the US since 2010\cite{yellowbook,Feroci2012,Feroci2014,Feroci2016}, comprises two instruments. The
Large Area Detector (LAD) consists of collimated arrays of silicon
drift detectors (SDDs) with a 1-degree field of view and a baseline
peak effective area of 10~m$^2$ at 8~keV (Fig.~\ref{fig:loft-p_effarea}), optimized for submillisecond
timing and spectroscopy of NSs and BHs. The sensitive Wide Field
Monitor (WFM) is a 2--50~keV coded-mask imager (also using SDDs) that
acts as a trigger for pointed LAD observations of X-ray transients and
also provides nearly continuous imaging of the X-ray sky with a large
instantaneous field of view.  

\begin{figure} [ht]
   \begin{center}
   \begin{tabular}{lr} 
   \includegraphics[height=7cm]{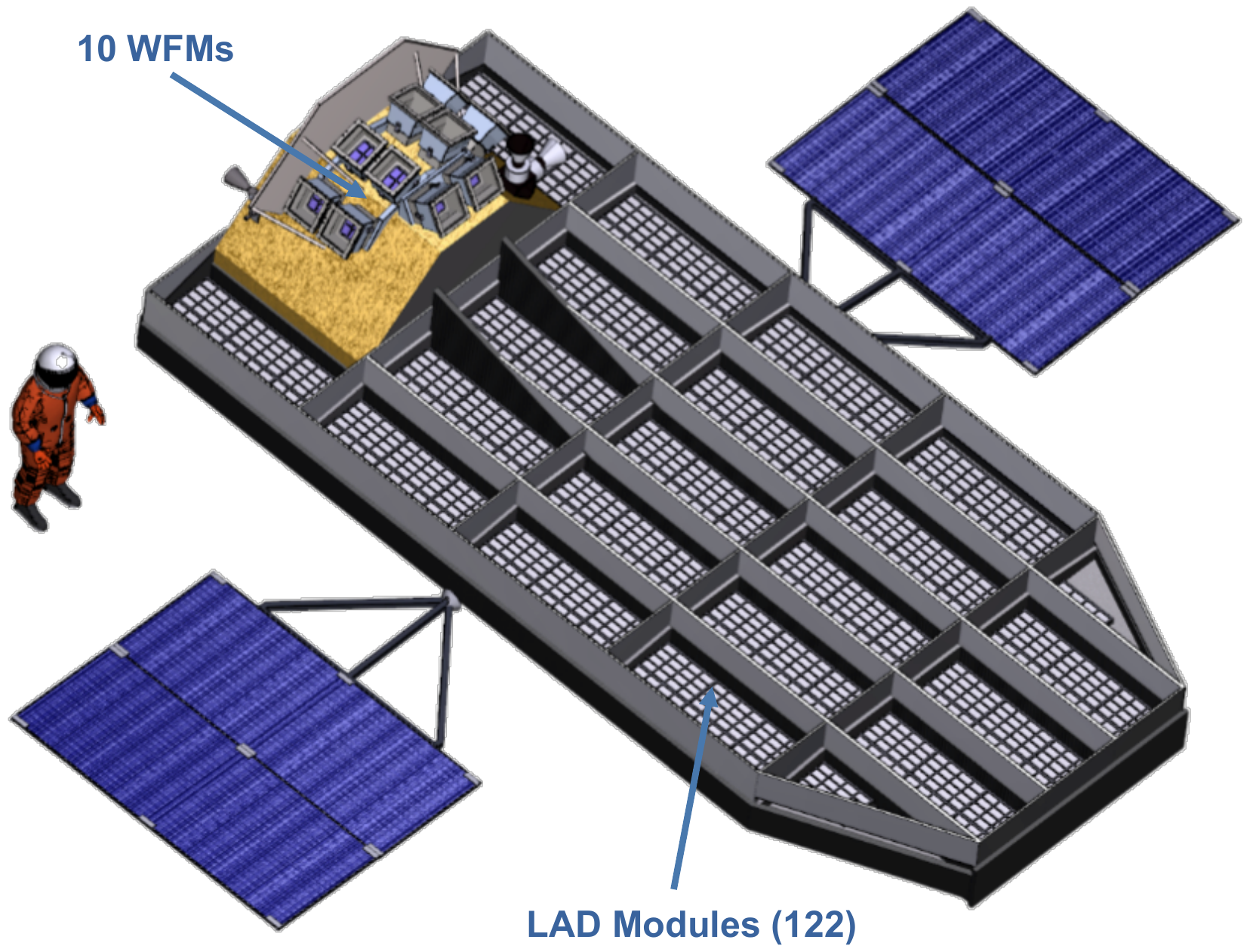} & \includegraphics[height=7cm]{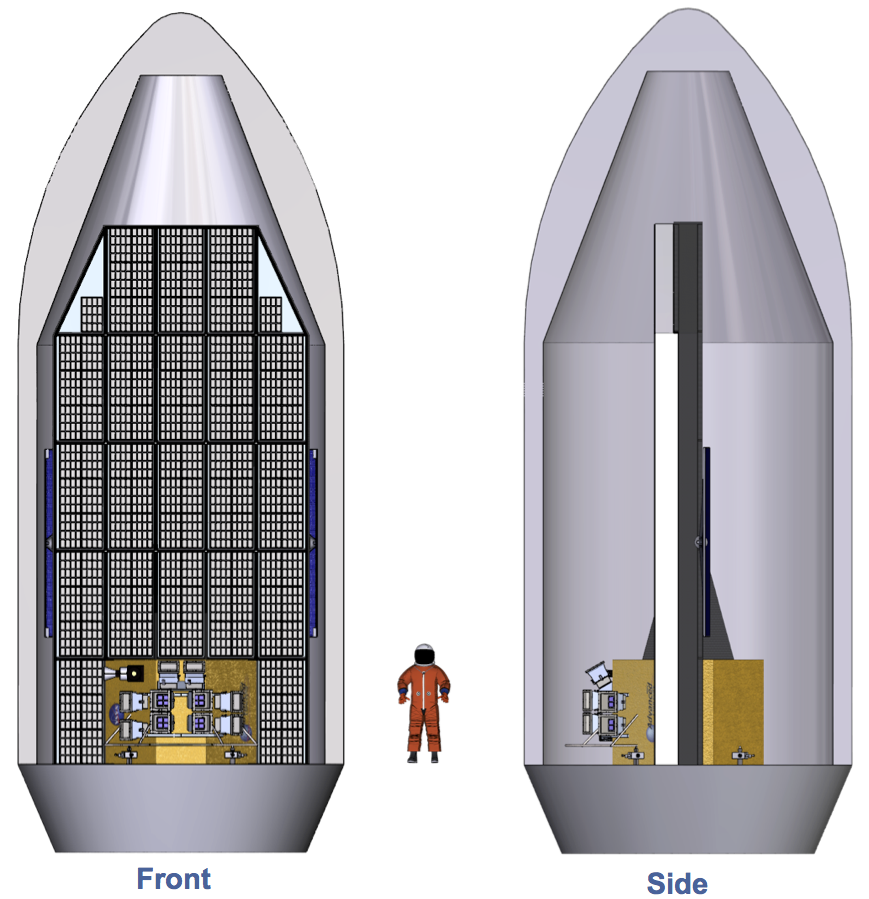}
   \end{tabular}
   \end{center}
   \caption{ \label{fig:loft-p_pic} LOFT-P spacecraft configuration with 122 LAD modules and 10 WFM cameras (left). This configuration fits within the volume of a Falcon 9 fairing (right). A Falcon Heavy is required to deliver LOFT-P to a 0 $\deg$ orbit from Cape Canaveral. An astronaut is added to both figures to give a sense of scale.}
   \end{figure}

We first presented LOFT-P as a concept, based on the ESA M3 studies of LOFT\cite{yellowbook}, at the American Astronomical Society (AAS) High Energy Astrophysics Division (HEAD): High-Energy Large- and Medium-class Space Missions in the 2020s meeting in 2015\footnote{\url{https://files.aas.org/head2015_workshop/HEAD_2015_Colleen_Wilson-Hodge.pdf}}, where it was well received. It was later presented as an example probe-class mission in the NASA Physics of the Cosmos Program Analysis Group (PhysPAG) final presentation to the head of NASA's Astrophysics Division, to demonstrate the strong community support for creation of a ``probe class,'' for NASA astrophysics missions that cost between \$500M and \$1B.  We submitted a white paper\cite{Wilson2016} describing LOFT-P science and this simple assessment to 
NASA's PhysPAG's Call for White Papers: Probe-class Mission Concepts, for which 14 white papers were received\footnote{\url{http://pcos.gsfc.nasa.gov/physpag/whitepapers.php}}. At the April 2016 HEAD meeting, NASA's PhysPAG endorsed the option that NASA issue a ROSES solicitation for Astrophysics Probe mission concept study proposals for input to the 2020 Astrophysics Decadal Survey\footnote{\url{http://pcos.gsfc.nasa.gov/physpag/meetings/head-apr2016/TH02_Bautz_HEAD_PCOS_update_Apr2016_v2.pdf}}. In May 2016 the Advanced Concepts Office at NASA MSFC performed a preliminary study (Fig.~\ref{fig:loft-p_pic}) to verify the cost of \textit{LOFT-P} as a US-led probe-class mission and to investigate a US-led design on a US launcher, in preparation.

\section{SCIENCE GOALS AND MISSION REQUIREMENTS}
\subsection{Science Goals}
{\bf Strong gravity and black hole spin.} Unlike the small
perturbations of Newtonian gravity found in the weak-field regime of
general relativity (GR), strong-field gravity results in gross
deviations from Newtonian physics and qualitatively new behavior for
motion near compact objects, including the existence of event horizons
and an innermost stable circular orbit (ISCO).  \textit{LOFT-P}
observations will probe strong gravitational fields of NSs and BHs in a way
that is complementary to gravitational wave interferometers like
LIGO and VIRGO. Accretion flows and the X-ray photons they emit are ``test
particles'' that probe the stationary spacetimes of compact objects,
whereas gravitational waves carry information about the dynamical
evolution of these spacetimes. As a result, \textit{LOFT-P} observations
will allow mapping the stationary spacetimes of black holes and
testing the no-hair theorem.  In GR, only two parameters (mass and
spin) are required to completely describe an astrophysical BH, and the
X-rays originating in the strong gravity regions necessarily encode
information about these fundamental parameters.  

\textit{LOFT-P} observations of accreting stellar-mass BHs will be
unique in providing three independent measurements of each BH spin
from high-frequency quasi-periodic oscillations (HFQPOs), relativistic
reflection modelling of Fe (and other) lines, and disk continuum
spectra, each using techniques with differing systematic uncertainties.  In
those systems in which HFQPOs have already been detected with $\sim
5$\% rms amplitude by \textit{RXTE}, deeper observations with
\textit{LOFT-P} will allow detections of the 5--10 additional QPO peaks
predicted by theory.  This will identify their frequencies with
particular linear or resonant accretion disk modes; this will be
possible once a spectrum of modes is observed, instead of just a pair.
\textit{LOFT-P}'s timing capabilities can also test whether the correct
spins have been obtained by reverberation mapping of the
X-ray reflection in X-ray binaries and AGN (for which it will
provide significantly better S/N than \textit{Athena}).

{\bf Properties of ultradense matter.} How does matter behave at the
very highest densities? This seemingly simple question has profound
consequences for quantum chromodynamics and for compact object
astrophysics. The equation of state (EOS) of 
ultradense matter (which relates density and pressure) is still poorly
known, and exotic new states of matter such as deconfined quarks or
color superconducting phases may emerge at the very high densities
that occur in NS interiors. This regime of supranuclear
density but low temperature is inaccessible to laboratory experiments
(where high densities can only be reached in very energetic heavy ion
collisions), but its properties are reflected in the mass-radius
($M$-$R$) relation of NSs. Consequently, measurement of NS $M$ and $R$
is the crucial ingredient for determining the ultradense matter EOS.

\textit{LOFT-P} will obtain $M$ and $R$ measurements by fitting
energy-resolved oscillation models to the millisecond X-ray pulsations
arising in a hot spot from rotating, accreting NSs. The detailed pulse
shape is distorted by gravitational self-lensing, relativistic Doppler
shifts, and beaming in a manner which encodes $M$ and $R$. Detailed
modeling of the pulse profile can extract $M$ and $R$ separately.
Measurements of both $M$ and $R$ for three or more NSs, made with
$\approx$5\% precision, would definitively determine the EOS of
ultradense matter, while measurement of a larger number of NSs with
$<$10\% precision would still place strong constraints.  The recently
approved \textit{NICER} mission will apply this same technique to
faint rotation-powered pulsars, a different class of NSs. This is
complementary to \textit{LOFT-P} rather than duplicative.  A key
difference between the NSs targeted by \textit{NICER} and
\textit{LOFT-P} is that the \textit{NICER} targets generally rotate
more slowly ($<$300 Hz) than the \textit{LOFT-P} targets ($>$600
Hz). As a result, \textit{NICER} observations cannot fully exploit
Doppler effects to break degeneracies between $M$ and $R$, making
precise and uncorrelated measurements more difficult.  \textit{NICER}
will obtain precise ($<$5\%) determinations of $R$ for only 1--2
sources; this is unlikely to be sufficient to solve the EOS problem,
since multiple measurements are required to measure the slope of the
$M$-$R$ curve and, hence, of the pressure-density relation that
describes the EOS.  Combining $M$-$R$ measurements from
\textit{NICER} and \textit{LOFT-P} will triple the sample size.

The $M$-$R$ relation of neutron stars can also be probed with
magnetar oscillations.  Like with the HFQPOs, by dramatically
improving the collecting area, enough frequencies should be found to
allow mode identification.  Given the relative precision of timing
calibration to response matrix calibration, timing-based models should
eventually allow the highest precision measurements possible.

{\bf Observatory science.} As a flexible observatory with superb
spectral-timing capabilities with wide field coverage of the sky over
a broad range of timescales, \textit{LOFT-P} will serve a large user
community and make significant scientific impact on many topics in
astrophysics. The LAD will study accretion physics, jet dynamics
(especially in conjunction with timing studies in the infrared that
will be possible on medium-sized telescopes), and disk winds (taking
advantage of the high throughput and high spectral resolution which
will allow very rapid detection of the turn-on of a disk wind).

The WFM's combination of angular resolution, sensitivity, spectral
resolution and {\it instantaneous} wide-field coverage will
enable studies of black hole transients, tidal disruption events, and
gamma-ray bursts too faint for current instrumentation.  It will also
enable instant spectroscopic follow-up of these events, as the
positional accuracy will be smaller than the fields of view of modern
integral field units.  The WFM's mission-long survey of the sky in Fe
K$\alpha$ will be more sensitive to Compton thick AGN than
\textit{eROSITA}.

The WFM will be unique as a discovery machine for the earliest stages
of supernova shock breakouts by working in the X-rays, and having the
sensitivity and instantaneous field of view to have an expected
detection rate of a few breakouts per year within 20 Mpc.  This will
allow much more rapid spectroscopic follow-up than other means of
discovering supernovae, allowing crucial studies of the early stages
of the explosions that can be used to probe details of the explosion
mechanisms and the binarity of supernova progenitors.  \textit{LOFT-P}
will be ideal for detecting and localizing X-ray counterparts to
gravitational wave sources, fast radio bursts, and optical transients
in the era of LSST.

\subsection{Mission Requirements}
For purposes of this LOFT-P study, the science requirements were assumed to be identical to those for LOFT M3\cite{yellowbook}. The large effective area and good spectral resolution were driven by the need to reduce Poisson noise for relatively bright sources to access weak timing features or to gather high-quality spectra for phenomena occurring on very short timescales. Examples include, simultaneously measuring both mass and radius for several neutron star systems to 3-5\%, directly observing millisecond orbital motion close to stellar mass black holes, and Fe-line tomography in AGNs to constrain the spin of the supermassive black hole. 

Because many of the target sources are highly variable, and because the desired observations can only occur in particular states, the Wide Field Monitor is required. Furthermore, the observatory needs to be relatively agile, able to respond to targets-of-opportunity in order to observe sources in the desired states and to respond to outbursts of new and interesting sources relevant to the LOFT-P science. 
\begin{table}
\caption{Instrument and Mission Requirements for LOFT-P\label{tab:instr_miss_req}}
\begin{minipage}{\textwidth}
\begin{tabular}{ll}
\hline
\hline
Parameter & Baseline \\
\hline
\hline
\multicolumn{2}{c}{\it Large Area Detector\cite{yellowbook}} \\
\hline
Effective Area & 9.5 m$^2$ @ 8 keV \\
Energy Range & 2--30 keV \\
Spectral Resolution & $<240$ eV @ 6 keV \\
Deadtime & $<$ 0.1\% @ 1 Crab \\
Time Resolution & 10 $\mu$s \\
Collimated field-of-view & 1$^\circ$ FWHM\\
Sensitivity (5 $\sigma$) & 0.1 mCrab in 100 s \\
\hline
\multicolumn{2}{c}{\it Wide Field Monitor\cite{yellowbook}} \\
\hline
Source Localization & 1 arcmin \\
Angular Resolution & 5 arcmin \\
Energy Range & 2-50 keV \\
Spectral Resolution & 300 eV @ 6 keV \\
Effective Area & 170 cm$^2$ (peak) \\
Field of view & 4.1 steradian \\
Sky Coverage & 50\% of LAD accessible sky \\
Sensitivity (5 $\sigma$) & 3 mCrab (50 ks) \\
\hline
\multicolumn{2}{c}{\it LOFT-P Mission} \\
\hline
Low Earth Orbit & 550 km, $< 5 ^\circ$ inclination \\
Sky visibility (Field-of-Regard) & $>$ 35\% ($>$ 50\% extended) \\
Pointing Accuracy & 1 arcmin on 3 axis \\
Pointing knowledge & 5 arcsec \\
Telemetry Rate & 100 Gbit/day \\
Slew Rate & 4$^\circ$/min \\
\hline
\end{tabular}
\end{minipage}
\end{table}

\section{SCIENCE INSTRUMENTS}

For purposes of the LOFT-P study, the science instruments were assumed to be identical to those described in the LOFT Yellowbook\cite{yellowbook}. They are described briefly below. Parameters of these instruments are listed in Table~\ref{tab:instr_miss_req}.
\subsection{Large Area Detector (LAD)}
The LAD provides the capability to revolutionize the study of X-ray variability on millisecond timescales. This instrument has previously been studied and described in detail\cite{Zane2014}. To provide that capability, two advances are needed over past instruments: dramatically larger area and improved spectral resolution. A modular design based on Silicon Drift Detectors (SDDs) is used to achieve the large area. Each LAD module has an array of $4 \times 4$ detectors and $4 \times 4$ collimators, the module back end electronics, and the ASICs, that control the detectors and read out the digitized events. For purposes of the LOFT-P study, the design unit is a LAD module, enabling the number of modules to be a study parameter. To meet the effective area requirement, 120 modules are needed. The field-of-view of the LAD is limited to 1 $\deg$ by lead glass micro-channel plate collimators. On the back side of each module, there will be a radiator for passive cooling and a shield to reduce the background. For every 25--30 LAD modules, there is a single panel back end electronics unit. Mass and power assumptions for the LAD modules for purposes of the LOFT-P study are listed in Table~\ref{tab:assumptions}. In the LOFT-P configuration, there are 122 LAD modules, vs. 126 in the LOFT M3 configuration\cite{yellowbook}.
\subsection{Wide Field Monitor (WFM)}
The WFM provides broad sky coverage to monitor potential LAD targets for transitions into desired observational states and provides considerable science in its own right. The WFM has previously been studeid and described in detail\cite{Brandt2014}. The WFM images the sky using coded mask cameras with solid-state class energy resolution, through the use of SDD detectors, the same detectors as are used for the LAD. Since the SDDs provide accurate positions in only one direction, pairs of orthogonal cameras are used to provide accurate source positions. The cameras have a Tungsten mask with a 25\% open area to optimize sensitivity for weaker sources. Like the LOFT M3 design\cite{yellowbook}, LOFT-P also includes 5 pairs of WFM cameras. Each camera pair has an effective field of view of $70^\circ \times 70^\circ$. Mass and power assumptions for the WFM are listed in Table~\ref{tab:assumptions}.

\section{MISSION DESIGN}
In this section, we describe the results of the LOFT-P mission concept study, a one month study performed by NASA MSFC's Advanced Concepts Office (ACO). MSFC's ACO is an engineering design facility for conceptual and preliminary design and analysis of launch vehicles, in-space vehicles and satellites, surface systems, human systems, and overall mission architecture concepts. The ACO is unique among the NASA preliminary design facilities because they have participated in development of every type of spacecraft flown by NASA. The team has the expertise to perform end-to-end analysis of new and innovative missions and vehicle systems. Details of ACO's capabilities, history, and people are provided here\footnote{\url{http://www.nasa.gov/centers/marshall/capabilities/adv_capabilities.html}}. Past and current studies of astrophysics missions include Hubble, Chandra, and X-ray Surveyor.  The goal of this study was to take a preliminary look at whether or not a US-led LOFT-P mission would fit within the \$500M-\$1B Probe class (excluding launch vehicle). 
\subsection{Assumptions and Requirements}
A new spacecraft was designed to meet the requirements listed in Tables~\ref{tab:instr_miss_req} and \ref{tab:assumptions}. Table~\ref{tab:assumptions} also lists key information about the science instruments based on the LOFT M3 study\cite{yellowbook}.

\begin{table}
\begin{center}
\caption{LOFT-P Mission Study Parameters\label{tab:assumptions}}
\begin{minipage}{\textwidth}
\begin{tabular}{ll}
\hline
\hline
Parameter & Required Value (Goal) \\
\hline
\hline
\multicolumn{2}{c}{\it LOFT-P Spacecraft/Mission} \\
\hline
\hline
Estimated Launch Year & 2027-2030 \\
Mission duration & 4 (5) years \\
Science data downlink & 6.7 Gbits (14 Gbits) per orbit \\
Orbit & LEO, Minimizing time in SAA, \\
 & 600 km upper limit, \\
 & $<5^\circ$ inclination \\
\hline
\multicolumn{2}{c}{\it LAD Module (each)\cite{yellowbook}} \\
\hline
Basic Mass & 6.05 kg \\
Power & 8.25 W \\
Thermal requirement & EoL LAD detector temperature \\
 & requirement of $-10^\circ$C over \\
 & nominal FoR; Up to $+11^\circ$ C for \\
 & extended FoR\\
Alignment & co-aligned within 3 arcmin\\
Quantity & 120 modules (minimum) \\
\hline
\multicolumn{2}{c}{\it Wide Field Monitor Camera (each)\cite{yellowbook}} \\
\hline
Basic Mass & 9.29 kg \\
Power & 7.56 W\\
Quantity & 10 cameras \\
\hline
\end{tabular}
\end{minipage}
\end{center}
\end{table}

\subsection{Mission Analysis}
The orbit selection was driven by minimizing three factors: passage through the South Atlantic Anomaly (SAA), radiation at higher altitudes, and atmospheric drag. 
Based on the ESA M3 study\cite{yellowbook}, an orbital inclination of $<5^\circ$ was required. To avoid higher radiation exposures, the altitude must be no greater than 600 km. Station-keeping requirements, driven mostly by atmospheric drag, determined the minimum altitude. The NASA Debris Assessment Software\cite{Opiela2007} (DAS) v2.0.2 was used to estimate the orbital altitude decay rate. Imposing a limit of 10 km degradation in altitude before raising the orbit back to the initial value resulted in a minimum recommended initial altitude of 550 km. Lower altitudes are possible, but station-keeping requirements increase substantially below 500 km, and do not offer any increase in payload mass capability to the launch vehicle. Analytical Graphics Systems Tool Kit (STK)\footnote{\url{http://www.agi.com}} was used to estimate ground station contact times, and determine the number of stations needed. Results indicated that in order to meet the daily science data download requirements, two ground stations are required. Details are provided in the Communications subsection below.
 
\subsection{Launch Vehicle}
Based on the ESA M3 study\cite{yellowbook}, a launch mass of 4070 kg was required. According to the NASA Launch Services Program (LSP) website\footnote{\url{http://www.nasa.gov/centers/kennedy/launchingrockets/index.html}} assuming launch from Cape Canaveral Air Force Station, the maximum launch mass that can be delivered by a Falcon 9 to a 500--600 km orbit with an inclination of  5$^\circ$ is 3705 kg, meaning that a Falcon 9 has insufficient performance to deliver LOFT-P to the required orbital altitude and inclination. NASA LSP stated that it is reasonable to assume that a Falcon Heavy or a similar vehicle will be on contract and available by the late 2020s. Since NASA LSP was not able to provide performance estimates, ACO estimated expected performance using the performance degradation from an SLS Block 1B going to a 28.5$^\circ$ vs a 0$^\circ$ orbit, and from the Falcon 9 to those same orbits. Applying this performance degradation to the advertised Falcon Heavy capability to 28.5$^\circ$ resulted in an estimated capability of 12,200 kg to an equatorial orbit. Applying the Falcon 9 performance degradation ratio resulted in an estimated worst case Falcon Heavy payload capability of 5630 kg to an equatorial orbit. Since launch vehicles with boosters lose less performance when going to lower inclinations, the study team feels that the 5630 kg estimate is much too conservative, with the actual capability being closer to the 12,200 kg estimate. Therefore, analysts used the average of the two values, and estimate a capability of 8900 kg to an equatorial orbit, easily placing LOFT-P into the desired location.

\subsection{Spacecraft Configuration}
The large payload dynamic envelope in the Falcon 9 fairing\footnote{\url{http://www.spacex.com/sites/spacex/files/falcon_9_users_guide_rev_2.0.pdf}} enabled a monolithic design for LOFT-P rather than the deployable design adopted for ESA M3 and M4 LOFT\cite{Feroci2012,Feroci2014}. This design accommodated 122 LAD modules, only slightly fewer than the 126 modules on LOFT M3, and the full 10-camera WFM, identical to LOFT M3. Additional WFM cameras could be added to the configuration if cost and telemetry allow it. The overall configuration is conservative and does allow room for component growth and for extra subsystem components to be added that were not analyzed in this study. Table~\ref{tab:mel} gives the master equipment list for this design. Masses include a 20\% mass growth allowance for structures, power, communication, command and data handling, guidance, navigation, and control, and for the science instruments. For the thermal control system, a 30\% mass growth allowance is included. For the propulsion system (excluding propellant) mass growth allowances of 5-25\% are used, depending on TRL and knowledge of specific components to be used.  Mass growth allowances are based on AIAA standards\footnote{\url{https://www.aiaa.org/StandardsDetail.aspx?id=3918}}. The total wet mass is the combined total of the spacecraft dry mass, science instruments, and propellant. 

\begin{table}
\caption{Master equipment list and mass budget for LOFT-P\label{tab:mel}}
\begin{minipage}{\textwidth}
\begin{tabular}{lr}
\hline
Equipment & Mass (kg) \\
\hline
Structures (incl. LAD frame panel) & 2160 \\
Thermal Control & 300 \\
Power & 190 \\
Avionics, Control, Comm. & 380 \\
Propulsion & 150 \\
Mass growth allowance\footnote{Per AIAA standards: 30\% for thermal control system; 7\% for \\
propulsion system (very high TRL); 20\% for all other systems}  & 530 \\ \hline
Spacecraft DRY MASS & 3710 \\
 &  \\
LAD (122 Modules) & 740 \\
WFM (10 cameras, ICU, \& harness) & 100 \\
Mass growth allowance (20\%) & 170 \\ \hline
Total Science Instrument Mass\footnote{Per ESA M3 Study} & 1010 \\
 & \\
Total Dry Mass & 4720 \\
Propellant (incl. 20\% mass growth allowance) & 940 \\ \hline
Total Wet Mass & 5660 \\ \hline
\end{tabular}
\end{minipage}
\end{table}

\subsubsection{Spacecraft Structure}
A finite element model was used to size the LOFT-P spacecraft and bus. MSC Patran was used to pre- and post-process the finite element model. MSC Nastran was used as the finite element model solver. Collier Research Hypersizer was used for the model optimization and sizing checks. Structural assessment includes strength, stability, and stiffness checks. Falcon Heavy envelope loads (launch/ascent) of 6 g axial and 2 g lateral were assesses. A constraint was applied at the LOFT-P Bus to payload adapter interface. The frame will be manufactured using Quasi-Isotropic IM-7 8552 composite laminates. This monolithic structure sizing is driven by stiffness. Structural deflections are well within the dynamic payload volume during launch and ascent. LAD mis-alignment due to non-uniform thermal loading can exceed the requirement of 3 arcmin if the thermal gradient is larger than $\sim 17 ^{\circ}$ C. The LOFT-P normal modes are low with first torsion at approximately 8 Hz.

\subsubsection{Communications System}
An X-band system with a fixed omnidirectional antenna is used for the downlink data system. A fixed antenna is more reliable and reduces mission risk as compared to a gimballed antenna.
A ground link analysis based on link times and daily accesses was performed to determine the best selection and number of required ground stations at 0, 5, and 10 degree inclinations. South Point Hawaii, Kourou, Guam, and Malindi were considered. South Point had no capability for a 0$^\circ$ orbit. Downlink averages were 3.8--5.6, 3.3--5.7, and 4.5--5.3 Gbits/orbit for 0, 5, and 10 degree, respectively, assuming a maximum X-band downlink rate of 10 Mbps, indicating that no single ground station gave sufficient time to download the required 6.7 Gbits/orbit of science data, and that 3--4 ground stations were required for the desired goal of 14 Gbits/orbit of science data. Initial investigations were started into using TDRSS, which allows downlink rates up to 300 Mbps, but requires a much higher power transmitter than is incorporated into the current LOFT-P design. Using TDRSS during launch and start-up operations is desirable, but further investigation into using TDRSS for normal operations is needed.

The communications system also includes a secondary VHF LOFT burst alert system with components based upon Orion EVA system heritage. This system will provide rapid alerts of transient events, e.g., gamma-ray bursts and X-ray bursts, to ground-based VHF receivers. 

\subsubsection{Power Systems}
The overall power demand, including the spacecraft, science instrumentation and 30\% mass growth allowance, is 2068W for the LOFT-P Falcon Heavy configuration. The power system supplies all of this demand. Power is generated by two conventional, folding, rigid panel solar arrays, 7.2 m$^2$ each, with a conversion efficiency of 25\% (beginning of life) and a total end of life power output of 3670W. The solar arrays were sized as folding rigid panel arrays using physics-based sizing relations based on manufacturers cell data. The power electronics sizing is based on flight heritage boards integrated into existing space qualified enclosures. Cabling is estimated using spacecraft dimensions and physics-based sizing tools. Cables are sized for a 2\% loss. Power requirements are aggregated from all other subsystems with a 30\% design margin per AIAA requirements. Energy storage is provided by six primary batteries. The power system mass (excluding mass growth allowance) is 191 kg.

\subsubsection{Avionics and GN\&C}
Two fully redundant Proton2x-Box flight computers from Space Micro are the core of the avionics system. These computers combine a commercial product set of building blocks, including a Proton400k processor, a power supply, DIO flash, up to 250 Gbit data storage, and 150 Mbs data rate transmission. 

Attitude knowledge is achieved using a redundant pair of Ball Aerospace star trackers and Northrop Grumman inertial measurement units (IMUs). The star trackers provide 4" of accuracy, meeting the 5" mission requirement. Both the star trackers and IMUs are at or above TRL 8. 

Pointing requirements for this spacecraft are modest, with a required pointing accuracy of 1 arcmin (3 $\sigma$) on 3-axis. Pointing stability is frequency dependent. The spacecraft will be normally inertially pointed, with uninterrupted observation times of about 1 ks to 100 ks (hours to days). Slew speeds will be about 2$^\circ$/min for normal slewing, with faster slews of 4$^\circ$/min for target-of-opportunity observations. For this study, an operational mode of slewing about the Y or Z axis was assumed. Because of the large mass and surface area of the system, damping launch tip-off rates is challenging, driving actuator sizing to unreasonably large sizes. Therefore, use of thrusters is recommended to damp tip-off rates. In our analysis, actuator sizing did not use tip-off rates.

Three axis drives were needed for 3-axis control, plus an additional one for single-fault tolerance. Control moment gyroscopes (CMGs) were selected because no reaction wheels were found that provided the required torque. The current design includes a pyramid configuration of 4 Ball Aerospace CMGs, with 129 Nms momentum storage, 2.64 Nm torque (up to 6.1 Nm as a set) to allow slew rates up to 4 $\deg$/min. A set of 3 Cayuga Astronautics L-series Magnetic torquers are used for continuous momentum unloading with 100\% margin, excluding tip-off rates.

\subsubsection{Thermal Control System}
Thermal control of the LOFT-P spacecraft will utilize passive high-TRL components such as MLI, white paint, passive radiators, and heaters to maintain spacecraft subsystem components within acceptable temperature ranges. A simplified model was developed in Thermal Desktop. The model was based on the LAD panel frame and spacecraft bus structures. A simplified thermal model of the LAD modules and front end electronics, based on the ESA M3 study\cite{yellowbook}, was incorporated into the LOFT-P thermal model. The analysis estimated the average temperature of the structural panel frame across the Field of Regard (FoR) and was used to size the thermal control components for the spacecraft. Hot and cold cases were studied with Sun beta angles for 0, 5, and 10 deg inclinations and 600 km orbits. Sun avoidance angles of 0 deg to 90 deg were also analyzed to evaluate the feasibility of meeting the LAD temperature requirements. A LAD detector temperature requirement of -10C, over the nominal FoR, is the driving requirement that influences thermal control. The FoR of the LAD constrains the solar flux seen by the LAD modules. The LOFT-P concept uses a local radiator design to lower the overall panel temperature without recourse to shading from sunlight. Analysis shows that the LAD structural panel average temperature is $<-10$C at a sun aspect angle of 30 deg, which compares well to the previous ESA designs. The LOFT-P concept provides additional conservatism due to the ability to shade the LAD modules with the primary LAD panel structure as well as mass margin for local sun shading if necessary. However, further analysis of the LAD modules and electronics needs to be performed to verify the overall thermal control approach. The WFM is protected from direct sunlight with a sun shield (as shown in Fig.~\ref{fig:loft-p_pic}) to avoid deformations of the coded mask\cite{Brandt2014}. The model was used to estimate the mass of a conceptual thermal control system for the spacecraft, propulsion system, and instruments. The estimated mass was 298 kg, not including 30\% mass growth allowance. Total estimated power of the thermal control system is 50 W.

\subsubsection{Propulsion System}
The propulsion system includes TRL 9+ hardware components and heritage derived hardware. The propulsion system's primary purpose is to de-orbit the spacecraft at the end of the mission, including 5 reentry maneuvers, and to perform orbit maintenance maneuvers. Secondary purposes include launch vehicle insertion error corrections, tip-off damping, collision avoidance, and momentum unloading. A simple monopropellant blowdown system with maximum off-the-shelf components, is selected for this task. The system consists of four PSI-ATK 80514-1 tanks that are loaded with hydrazine and nitrogen pressurant. The thruster configuration comprises 4 pods, each containing three Aerojet MR-104 attitude control system thrusters (2N). One pod also includes an orbit adjust thruster (440 N), an Aerojet MR-111E. The system is single fault tolerant at the component level, two fault tolerant to failure at the system level. The system provides a total delta-V (with margin) of 298 m/s. Margins are 25\% for launch vehicle insertion errors, orbit maintenance, collision avoidance, and momentum unloading. For reentry, for which the delta-V is well determined, a 10\% margin is assumed. Tank sizing allows up to 378 m/s delta-V. The predicted dry mass of the system without contigency is 154 kg. Margins are low (5\%) for the high-TRL off-the-shelf components such as the hydrazine tanks, the thrusters, and the isolation latch valve. Propellant dominates the mass of the system, with 894 kg of hydrazine and 46 kg of nitrogen pressurant including 20\% mass growth allowance.  

\subsubsection{Preliminary Cost Estimate}
Costs for the LOFT-P mission were estimated using the following parametric models PCEC (Project Cost Estimating Capability), SEER-H, NICM (NASA Instrument Cost Model), and MOCET (Mission Operations Cost Estimating Tool) for ground data systems/mission operations systems costs. Two cost estimates were performed during the study. The first was based on the ESA M3 study of LOFT\cite{yellowbook}. The second was based on the MSFC Advanced Concepts Office study of LOFT-P. Both were assumed to be NASA-led for cost assumptions. The NASA Standard Level WBS for space flight projects was assumed, based on NPR 71020.5E: NASA Space Flight Program and Project Management Requirements, Appendix H. Costs were estimated in FY2016 dollars, with a fee of 12.5\%, and cost reserves at 35\%. Launch vehicle costs were excluded from both estimates. Mass with contingencies was used. MOCET was used to calculate all phase E costs, based on the Fermi mission,

The costs for both concepts assumed the following for the LAD: 125 detector modules, 5 Panel Back end Electronics, 2 ICUs. Costs were based on one development and production of 125 modules. For the WFM, the model assumed 10 cameras, 2 WFM ICUs. Costs were based on one development and production of 10 cameras. Average modification on the electronic components of both instruments was assumed, given the considerable development that has already taken place in Europe. For both concepts, major modification for the spacecraft structure, average modification for the C\&DH system, and 
minor modification for the electrical, thermal, propulsion, and communication systems were assumed. The same phase A-D schedules taken from\cite{yellowbook} were used. For consistency with ESA estimates, a 3 year mission (5 year goal) was assumed for LOFT M3, while for the LOFT-P mission was assumed to have a duration of 4 years (5 year goal). Using these cost models, our preliminary cost estimates show a 15-25\% margin with respect to the \$1B Probe-class cost cap, including 35\% cost reserve. Both cost estimates include full life cycle costs, including labor, instruments, spacecraft, mission operations, and ground data systems. Both cost estimates compare well with other astrophysics missions in the ONCE database, including Fermi.    

\section{FUTURE WORK \& CONCLUSIONS}
The LOFT-P concept complements existing LOFT designs and bounds options. A single panel was chosen for LOFT-P to reduce complexity, but requires increased mass to meet stiffness and stability requirements. The large panel manufacturing and mass may offset the reduced complexity. Future studies need to trade a single LAD panel vs a multipanel deployed configuration, including analysis for low frequency vibrations, to verify impacts from LAD module assembly and alignment, and to assess the impacts on overall spacecraft maneuverability and stability. The large moment of the monolithic design drives the need for thrusters to control tip-off and the need for CMGs, which limits fast slew rates and would likely be a major driver in a future trade study of a single panel vs multipanel design. Further studies are also needed for the fast slew rate, including considering feasibility of using thrusters for fast slews, which would likely allow for the use of reaction wheels instead of the more-expensive CMGs for attitude control.  Cost fidelity can also be improved by refining the mass basis and investigating instrument/component modeling, including definition of heritage/high TRL components for model inputs and conducting a sensitivity analysis.

The LOFT-P study has shown that a LOFT-like mission is feasible as a probe-class mission. The estimated cost of the monolithic LOFT-P design is similar to the multipanel LOFT M3 design. This study has positioned LOFT-P well for a more detailed concept study in preparation for the 2020 Astrophysics Decadal Survey. LOFT-P science is timely. With its highly capable LAD and WFM, LOFT-P will address fundamental physics, and time-domain science.

\acknowledgments 
 NRL`s work on X-ray astrophysics is funded by the Chief of Naval Research (CNR). The LOFT-P study was funded internally by NASA MSFC. The work of the MSSL-UCL and Leichester SRC on the LOFT-LAD project has been supported by the UK Space Agency. The work of the ICE (CSIC-IEEC) on the LOFT-WFM project has been supported by funds from the Spanish MINECO.
\bibliography{report} 
\bibliographystyle{spiebib} 

\end{document}